# Generative AI in Knowledge Work: Perception, Usefulness, and Acceptance of Microsoft 365 Copilot

Carsten F. Schmidt[1], Sophie Petzolt[1], Wolfgang Beinhauer[1], Ingo Weber[2,3], Stefan Langer[2]

[1]Fraunhofer Institute for Industrial Engineering IAO, Stuttgart, Germany; [2]Fraunhofer-Gesellschaft zur Förderung der angewandten Forschung e.V., Munich, Germany; [3]Technical University of Munich, School of CIT, Munich, Germany

Corresponding author: Carsten F. Schmidt (carsten.schmidt@iao.fraunhofer.de)

**Keywords**

**Abstract**

The study analyzes the introduction of Microsoft 365 Copilot in a non-university research organization using a repeated cross-sectional employee survey. We assess usefulness, ease of use, output quality and reliability, and usefulness for typical knowledge-work activities. Administrative staff report higher usefulness and reliability, whereas scientific staff develop more positive assessments over time, especially regarding productivity and workload reduction. Copilot is widely viewed as user-friendly and technically reliable, with greatest added value for clearly structured, text-based tasks. The findings highlight learning and routinization effects when embedding generative AI into work processes and stress the need for context-sensitive implementation, role-specific training and governance to foster sustainable acceptance of generative AI in knowledge-intensive organizations.

**Practical relevance**

The study provides empirical insights into the implementation of generative AI in knowledge-intensive organizations. It shows that Microsoft 365 Copilot particularly enhances efficiency in administrative areas, while context-specific training and support measures are crucial for success in research settings. The findings help organizations design context-sensitive AI implementation strategies, foster user acceptance, and purposefully leverage the potential of AI to ensure productive human–AI interaction and create organizational value.

# 1 Introduction

Today's world of work is undergoing a profound technological transformation, driven in particular by the capabilities of artificial intelligence (AI). AI technologies can be defined as "a collection of interconnected technologies used to solve problems and perform tasks that, when performed by humans, require thinking" (Bankins et al., 2024). The potential impacts of generative AI models, which represent a subset of AI technologies (Saif et al., 2024), on the economy, society, democracy, and the environment are being debated intensively (Budhwar et al., 2023). Tools based on large language models such as ChatGPT or Microsoft 365 Copilot (M365 Copilot) have attracted considerable attention due to their ability to generate text, analyze inputs, and provide context-appropriate completions (Saif et al., 2024). At the same time, their introduction and use have triggered controversial debates about implications for employees and jobs (Budhwar et al., 2023; Frey & Osborne, 2017).

The integration of generative AI into the economy and academic research is already receiving broad scholarly and media attention. However, the use of such tools in non-university research organizations (German: außeruniversitäre Forschungseinrichtungen, AUF) has so far been insufficiently researched. At the same time, such organizations play a central role in national innovation ecosystems (OECD, 2011). As intermediary organizations between academic research and industrial application, they serve as a bridge within the innovation system. They combine scientific excellence with a strong orientation toward application and operate at the interface of basic research, technology development, and economic exploitation. This multiple role creates specific organizational requirements as well as particular opportunities when deploying new technologies such as generative AI. In Germany in particular, non-university research organizations contribute substantially to innovation capacity and global competitiveness by generating knowledge that feeds into private-sector innovation (Graf & Menter, 2022).

Against this background, the present article aims to examine the introduction and use of M365 Copilot in a non-university research organization, differentiated by scientific and administrative areas of work. The following research questions are central:

F1: How do scientific and administrative employees differ in their assessment of the key acceptance constructs of M365 Copilot (perceived usefulness, perceived ease of use, output quality, and reliability)?

F2: How do these assessments change between the measurement points T01 and T02 in both employee groups?

F3: For which typical knowledge-work activities is M365 Copilot perceived as particularly useful or less useful by scientific and administrative employees?

The study is exploratory and aims to derive initial empirical patterns and hypotheses regarding the task-related relevance of generative AI tools along these questions.

## 2 Theory and related research

The capabilities of artificial intelligence to perform human-like work have improved rapidly in recent years (Dell'Acqua et al., 2023). It is increasingly apparent that AI capabilities overlap with those of humans. Consequently, integrating and combining human work with AI creates new fundamental challenges and opportunities for organizations, as it changes the future of work especially in knowledge-intensive domains (Bankins et al., 2024; Dell'Acqua et al., 2023; Nejjar et al., 2024).

The introduction of AI in organizations is discussed controversially. While some studies point to potential job losses (Frey & Osborne, 2017), other studies emphasize possible productivity and quality gains (Jarrahi, 2018). Recent analyses further suggest that, in contrast to earlier waves of automation, AI adoption particularly affects higher-paid jobs and knowledge-intensive activities (Chugunova et al., 2025; Eloundou et al., 2024). The literature largely agrees that AI tools can offer substantial productivity advantages (Brynjolfsson et al., 2025; Dell'Acqua et al., 2023; Peng et al., 2023; Chugunova et al., 2025). At the same time, current studies emphasize that although AI tools can increase productivity, their effects depend crucially on how the technology is introduced and on users' skills (Chugunova et al., 2025).

While prior research has focused strongly on the introduction of generative AI in companies (e.g., Singh et al., 2024) and on the university context (Intarakumnerd & Goto, 2018), implementing generative AI is particularly relevant for non-university research organizations, which often face increased innovation pressure and efficiency demands (Meier, 2024). Generative AI opens new possibilities for scientific work and knowledge transfer. Generative AI, driven by large language models (LLMs), can support researchers in solving complex tasks and developing innovative approaches (Intarakumnerd & Goto, 2018). For example, generative AI can support the drafting of research manuscripts, code development, and experiment design (Nejjar et al., 2024), thereby influencing interactions within and with organizations (Deloitte, 2025; Gupta et al., 2023). Despite this growing interest in the introduction and use of generative AI across sectors, there remains a significant research gap regarding its specific added value and practical application in non-university research organizations. In particular, empirical evidence is lacking on which tasks are supported by generative AI, how the technology is integrated into everyday work processes, and what effects it has on organizational efficiency and individual productivity.

Work in non-university research organizations includes experimental research, the development and testing of prototypes, and continuous data collection. In addition to scientific work, project management is of central importance: researchers often assume tasks such as project planning, acquisition of funds, and budget control—activities that are typically more pronounced than in universities.

Accordingly, scientific employees also engage to some extent in administrative tasks. Knowledge transfer takes place via specialist publications, conferences, and the economic as well as political exploitation of research results. Beyond scientific work, extensive administrative tasks also arise in the Fraunhofer Society—from payroll and financial controlling to facility management. The Fraunhofer headquarters sets strategic and administrative frameworks, while the approximately 70 institutes operate autonomously within this framework and manage their administrative processes operationally. Compliance and legal issues are particularly salient. Data protection as well as patent and copyright law must be complied with precisely. In this context, both correctness and efficiency in task execution are decisive.

These theoretical considerations can be integrated into an analytical framework that focuses on perceived usefulness, perceived ease of use, output quality, and reliability of M365 Copilot as key dimensions of acceptance and links them to task-related usefulness assessments.

On this basis, the analyses explicitly focus on (a) group-specific differences between scientific and administrative areas of work (F1), (b) temporal changes in perceived acceptance characteristics during the early implementation period (F2), and (c) task-related patterns of the usefulness of M365 Copilot for different forms of knowledge work (F3).

# 3 Method and sample

The present study examines the extent to which employees of a large non-university research organization perceive the benefits of M365 Copilot. Particular attention is paid to possible differences between administrative and scientific areas of work.

Because only limited hypotheses are available to date regarding the task-related relevance of generative AI tools, a quantitative, inductive research approach was chosen in order to derive initial insights and hypotheses.

## 3.1 Sample and recruitment

The study was conducted as part of the technical pilot of M365 Copilot in the non-university research organization (German: AUF). In May 2024, 550 employees were equipped with a corresponding license as part of a quota sample and were simultaneously invited to participate in an accompanying survey. The selection was aligned with the overall organizational structure with the aim of achieving a depiction of the workforce that is as realistic as possible with respect to institute affiliation, age groups, genders, areas of work, and differing levels of experience in dealing with artificial intelligence. In coordination with the institutes and central units, persons were nominated for this purpose; thus, no formal random sample was drawn. Recruitment for the survey took place via a centralized invitation by email.

Data collection was multi-stage: first, profile data were collected, followed by recurring surveys on usage situations and perceived benefits. These initially took place on a biweekly basis and later every four weeks. For the present analysis, two survey periods were considered:

T01 (November–December 2024) with a net sample of 106 persons (40 from administration and 66 from the scientific area),

T02 (March–April 2025) with 39 participants from administration and 51 from the scientific area, totaling 90 persons.

All persons with an M365 Copilot license in the organization were invited in each period. For technical implementation, pseudonymized IDs were assigned so that, in principle, individual persons could be matched across the two measurement points.

However, the actual overlap of participants who took part in both T01 and T02 is relatively small. For this reason, the data are treated analytically as two repeated cross-sectional samples. The differences between T01 and T02 reported below therefore refer to changes at the sample level and allow only limited statements about individual trajectories.

This approach allows initial statements and hypotheses about differences between the groups, but it does not replace a representative study. For valid statements about the target population of around 32,000 employees, approximately 380 valid cases would be required; given 550 licensed persons and an expected response rate of about 20%, the study's inferential power for the overall population is accordingly limited.

## 3.2 Survey instrument and procedure

The study is methodologically an exploratory quantitative study with deductively selected, theory-guided constructs. In contrast to a strict model test (e.g., of a full TAM or UTAUT), the constructs collected are primarily analyzed descriptively in order to identify patterns and differences between employee groups and measurement points in an exploratory manner and to derive hypotheses for further research. Data were collected using a standardized online questionnaire created with LimeSurvey.

To capture attitudes and usage experiences, established constructs from acceptance research are used (including perceived usefulness, perceived ease of use, output quality, reliability), which are aligned with classic technology acceptance models (e.g., Davis, 1989). The basis comprises validated scales from the Technology Acceptance Model (TAM) by Davis (1989) and its extension by Kohnke (2015), in particular for perceived usefulness and perceived ease of use, supplemented by task-related dimensions according to Reinhardt et al. (2011). However, a specific acceptance model (TAM, TAM3, UTAUT) is not tested in the strict sense in this article.

The questionnaire consisted of several thematic blocks: (a) socio-demographic and work-related information: age, gender identity, educational attainment, area of work (scientific or administrative), professional experience, institute, research field, and current occupational group; (b) AI competence: type of knowledge acquisition regarding M365 Copilot (e.g., trainings, self-study, exchange with colleagues); (c) attitudes and experiences regarding use: including output quality (1 item), voluntariness of use (1 item), perceived usefulness (5 items), perceived ease of use (2 items), reliability (1 item), and application-specific usefulness (9 items). The operationalization of the variables is directly aligned with the guiding research questions.

All items were measured on a seven-point Likert scale ("strongly disagree" to "strongly agree"). For analysis, responses were coded numerically (−3 = "strongly disagree" to +3 = "strongly agree").

In order to enable honest and unfiltered feedback—especially against the background of controversial public debates about the use of generative AI (e.g., Ball, 2023; Wang et al., 2023; Chugunova et al., 2025)—the survey was conducted in a pseudonymized manner. The design also aimed to reach participants who made only limited use of M365 Copilot despite having trial access or who generally had less experience with generative AI. Nonetheless, a certain self-selection effect in favor of more interested or technology-affine persons cannot be ruled out.

Respondents' information is based on subjective assessments and individual usage experiences. It therefore does not represent an objective measurement of the effectiveness or impacts of AI use and is subject to the usual limitations of self-reported data (e.g., recall bias, social desirability).

Within the survey period from November 2024 to April 2025, a total of eight clearly dated update cycles for M365 Copilot were documented. These updates each marked further developments and adjustments of underlying functions and models, which in some cases also entailed noticeable changes in the user interface and in the system's response behavior. They included both content-related extensions (e.g., new integrations in Office applications and improved prompt interpretation) and technical optimizations in the areas of system stability, reliability, and security (Microsoft Corporation, 2025).

### 3.3 Data preparation and analysis

To handle missing values, a threshold procedure was applied. Cases with more than 20% missing responses across all relevant items were fully excluded from the dataset. Cases with a lower proportion of missing values were retained in the analyses; remaining missing values were not imputed but were excluded listwise depending on the analysis. In addition, datasets with obviously implausible response patterns (e.g., completely identical responses across all items) were checked and excluded from the analyses.

Analyses proceeded in several steps. First, descriptive statistics (means, standard deviations, sample sizes) were computed separately by area of work (administration vs. science) and measurement point (T01, T02). For inferential assessment of group and time differences in key acceptance constructs (perceived usefulness, perceived ease of use, output quality, reliability), independent-samples t-tests assuming unequal variances (Welch correction) were performed. Effect sizes (Cohen's d) were also reported.

Task-related usefulness assessments were captured conceptually as separate dimensions and analyzed in an exploratory descriptive manner. Given the large number of potential single comparisons and the associated inflation of the alpha error (increased probability of false-positive findings in multiple tests), inferential single-item tests were not conducted for these items. The results primarily serve pattern recognition and hypothesis generation.

Because the data consist of two repeated cross-sectional samples, time comparisons refer to differences at the sample level and do not allow statements about intra-individual change processes.

### 3.4 Scale quality

To examine internal consistency of the five perceived usefulness (PU) items, Cronbach's alpha was calculated separately for both measurement points. The scale shows very high internal consistency (T01: α = .97; T02: α = .95). Corrected item–total correlations ranged between .91 and .96 (T01) and between .85 and .93 (T02). All items were therefore retained in the scale.

The two-item scale for perceived ease of use showed stable inter-item correlations (T01: r = .58; T02: r = .66), indicating consistent measurement of the construct. For two-item scales, the inter-item correlation is the key measure of internal consistency.

The constructs output quality, reliability, harm from incorrect output, and voluntariness were each measured as single items; internal consistency cannot be calculated methodologically for these. The same applies to the task-related usefulness assessments, which were operationalized as distinct task dimensions and were not aggregated into a common index.

## 4 Results

In the following chapter, the results of the survey on the use and perception of M365 Copilot in a non-university research organization are presented. The aim of the analyses is to answer the research questions F1 to F3 formulated in Section 1.

Section 4.1 primarily addresses F1 and F2 by examining group-specific assessments of the key acceptance dimensions and their development between T01 and T02. Section 4.2 additionally focuses on F3 by analyzing the perceived usefulness of M365 Copilot for different knowledge-work activities.

### 4.1 Perceptions of output quality, reliability, ease of use, and usefulness

Figure 1 illustrates how the surveyed constructs develop over time and provides an overview of perceived usefulness, perceived ease of use, output quality, and reliability of M365 Copilot from the perspective of the two employee groups. The results are presented in detail below, and developments over time for both groups are discussed.

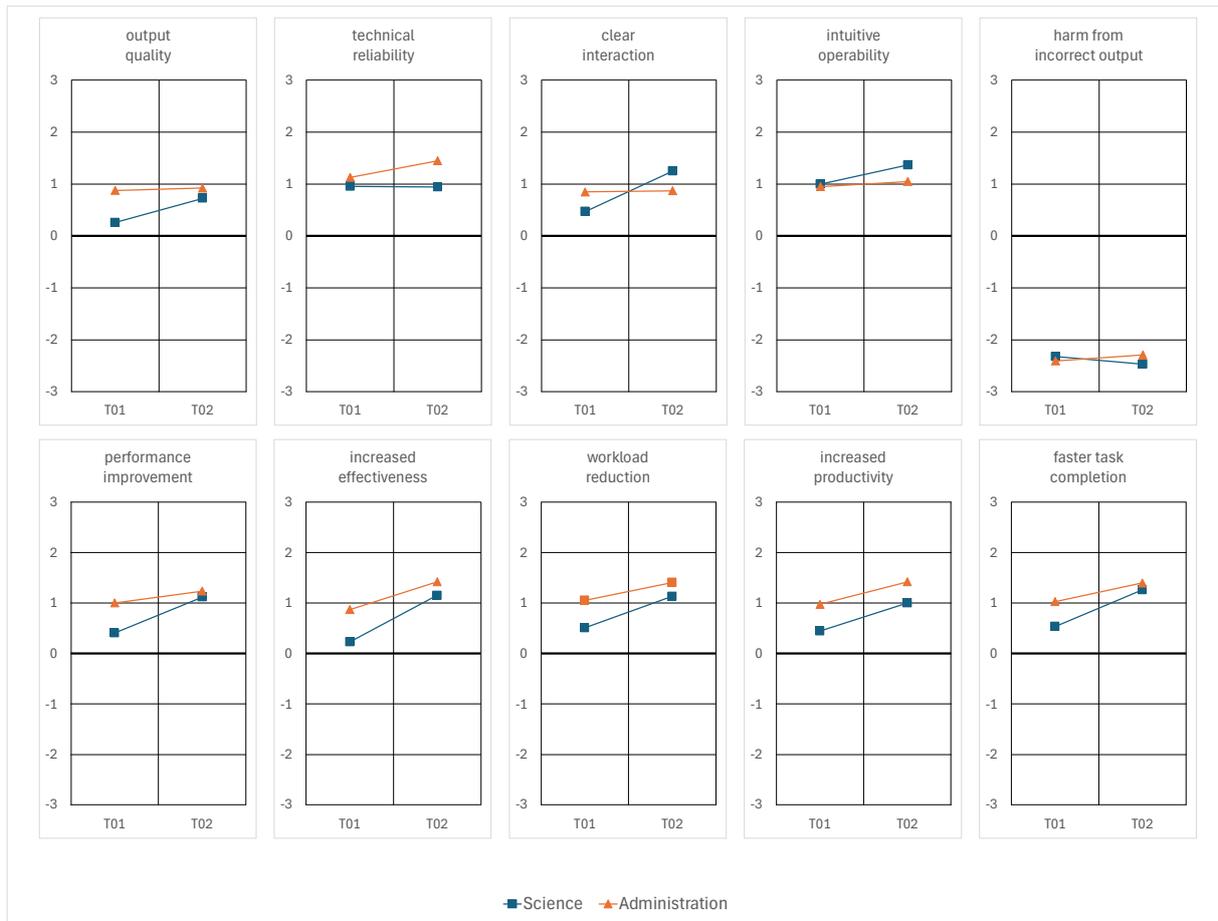

**Figure 1.** Comparison of perceived usage and benefit aspects of M365 Copilot in the employee groups "Science" and "Administration" at T01 and T02. The figure shows group-specific means of perception-based assessments on a scale from –3 ("strongly disagree") to +3 ("strongly agree").

To statistically substantiate the mean differences shown in Figure 1, independent Welch t-tests were conducted. Table 1 summarizes group and time comparisons including effect sizes (Cohen's d) and two-sided p-values.

Table 1. Means (M), effect sizes (Cohen's d), and p-values for group and time comparisons of key acceptance constructs.

| Construct | Group | T01 M | T02 M | d (time) | p (time) | d (group) T01 | p (T01) | d (group) T02 | p (T02) |
|---|---|---|---|---|---|---|---|---|---|
| Perceived usefulness | Scientific | 0.42 | 1.09 | 0.42 | .022 | 0.39 | .050 | 0.20 | .332 |
|  | Administrative | 0.94 | 1.35 | 0.38 | .092 | — | — | — | — |
| Perceived ease of use | Scientific | 0.74 | 1.31 | 0.45 | .016 | 0.13 | .497 | 0.27 | .209 |
|  | Administrative | 0.90 | 0.98 | 0.06 | .781 | — | — | — | — |
| Output quality | Scientific | 0.26 | 0.73 | 0.34 | .067 | 0.48 | .018 | 0.17 | .440 |
|  | Administrative | 0.88 | 0.92 | 0.04 | .853 | — | — | — | — |
| Reliability | Scientific | 0.96 | 0.94 | 0.01 | .955 | 0.12 | .557 | 0.34 | .114 |
|  | Administrative | 1.13 | 1.15 | 0.24 | .282 | — | — | — | — |
| Harm from incorrect output | Scientific | -2.32 | -2.47 | 0.11 | .556 | 0.08 | .684 | 0.15 | .491 |
|  | Administrative | -2.41 | -2.29 | 0.12 | .594 | — | — | — | — |
| Voluntariness of use | Scientific | 2.63 | 2.79 | 0.23 | .204 | 0.26 | .192 | 0.13 | .541 |
|  | Administrative | 2.80 | 2.69 | 0.15 | .516 | — | — | — | — |

**Perceived output quality is positive in both groups at both measurement points.**

The output-quality question captures respondents' assessment of the generated content with respect to correctness, completeness, and overall quality. In both groups, mean values at T01 and T02 are above the midpoint of the scale.

Scientific employees rate output quality at T01 lower than administrative employees. Between T01 and T02, scientific employees show an increase that does not reach the significance threshold. For administrative employees, no significant time difference emerges. In the group comparison, there is a significant difference in favor of administration at T01, which is no longer statistically significant at T02.

**The survey captures the assessment of stable and error-free functioning of M365 Copilot.**

In the survey, the technical reliability of M365 Copilot—understood as stable and error-free functioning—was assessed.

Ratings in both groups at both measurement points are in the positive range. For administrative employees, a slight but non-significant increase over time is observed. For scientific employees, there is likewise no significant time difference.

Overall, there are neither statistically meaningful changes over time nor significant group differences regarding perceived reliability.

**Perceived ease of use improves over time, particularly among scientific employees.**

Perceived ease of use of M365 Copilot captures the subjective assessment of the system's comprehensibility, transparency, and intuitive operability. It reflects how effortless interaction with Copilot is from users' perspective.

Ratings in both groups at both measurement points are in the positive range (see Table 1). For scientific employees, there is a significant increase in perceived ease of use between T01 and T02. For administrative employees, no significant time difference is found.

At neither measurement point are there significant group differences. Overall, the pattern indicates that scientific employees in particular perceive the interaction as more comprehensible over time, whereas assessments in administration remain stable.

**Assessments of potential harm from incorrect output remain stably low.**

The item "harm from incorrect output" captures the perceived extent of potential harm due to erroneous outputs or false statements by M365 Copilot.

Ratings in both groups at both measurement points are in the negative range of the scale, indicating a low perceived frequency of harm (see Table 1).

No statistically significant differences are found either for within-group time comparisons or for group comparisons between science and administration.

**Both groups report a positive assessment of performance-related effects of M365 use, with a stronger increase among researchers.**

The five items on performance improvement, increased effectiveness, workload reduction, increased productivity, and faster task completion jointly form the perceived usefulness (PU) construct in the sense of the Technology Acceptance Model (Davis, 1989). They capture the extent to which users perceive the use of M365 Copilot as performance-enhancing, efficiency-increasing, and workload-reducing.

In both groups, ratings at both measurement points are in the positive range. For scientific employees, there is a significant increase in perceived usefulness between T01 and T02. Among administrative employees, a positive but non-significant increase is also observed.

In the group comparison, there is a significant difference in favor of administration at T01. At T02, this difference is no longer statistically significant. Overall, the pattern suggests that usefulness attributions increase over time particularly among scientific employees, whereas assessments in administration remain at a stably positive level.

In substantive terms, these assessments concern both perceived increases in effectiveness and productivity as well as workload reduction and faster task completion. The results thus show that M365 Copilot is perceived as supportive for task completion across groups, with assessments converging over time.

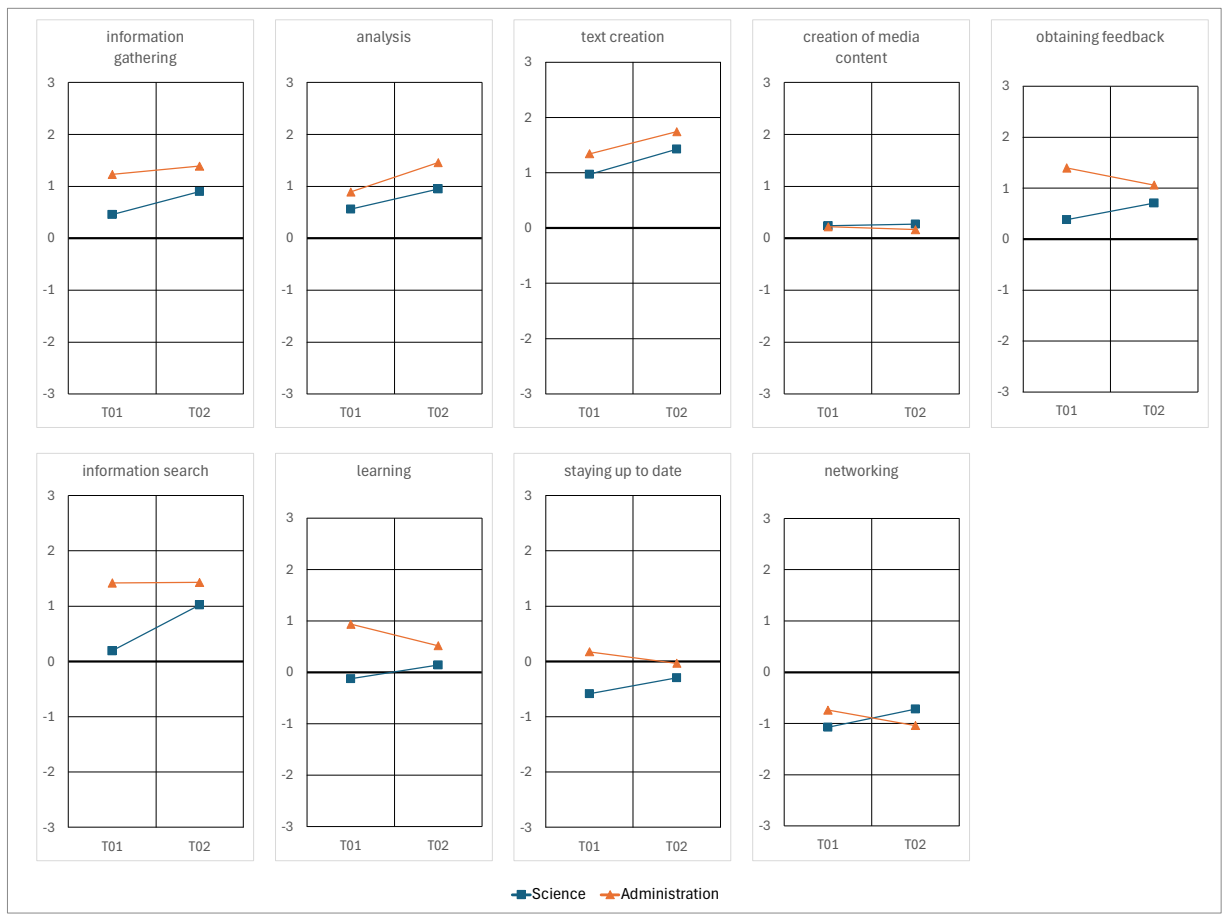

**Figure 2.** Comparison of the task-related usefulness of M365 Copilot in the employee groups "Science" and "Administration" at T01 and T02. The figure shows group-specific means of perception-based assessments on a scale from –3 ("strongly disagree") to +3 ("strongly agree").

### 4.2 Task-related usefulness of M365 Copilot

In Figure 2, the development of perceived usefulness of M365 Copilot for completing various typical knowledge-work activities is shown. The results are then reported in detail and the temporal development for both groups is traced systematically.

The pattern already visible in Figure 1 is clearly confirmed: administrative employees consistently rate the usefulness of M365 Copilot higher than employees in the scientific area. At the same time, a slight to moderate increase in perceived usefulness between T01 and T02 can be observed in both groups—an indication of growing familiarity with the system or improved integration into existing workflows.

Activities such as information gathering, analysis, and text creation are evaluated particularly positively. The administrative group remains consistently in the positive range and, in some cases, shows further increases over time. A positive trend is also apparent in the scientific group, albeit at a lower overall level. This suggests that Copilot is experienced as helpful especially where tasks are clearly structured, automatable, and text-based.

A more differentiated picture emerges for the creation of media content: both groups are in the neutral range here, with only minor changes over time. This indicates that support in this area is either perceived as technically still insufficient or as less relevant for the respective work context.

Clearer differences emerge for obtaining feedback: while administrative employees rate this aspect clearly positively and slightly increase their assessment, the rating in the scientific group even declines over time. This development suggests that Copilot is used more strongly in administration for text- and communication-related tasks, whereas benefits in the scientific context—especially for complex or research-related feedback—are perceived as limited.

In the areas of information search and learning, a similar pattern to text creation emerges: the administrative group remains at a high level with slight gains, while the scientific group remains more reserved overall but also shows an upward tendency. Particularly in the area of learning, ratings remain rather muted—an indication that Copilot is understood less as a learning tool and more as operational support for tasks.

In the categories "staying up to date" and "networking," assessments are overall negative or neutral. Both groups see little added value here; for networking, perceptions are even slightly negative. Apparently, these activities are either not associated with the current functions of the system or the available functions are perceived as insufficient for the requirements of social or informal interactions.

The task-related assessments can be grouped into three patterns:

- 1) clearly positive ratings for structured, text- and information-related tasks,
- 2) mostly neutral ratings for media-related or creative activities,
- 3) little support for social and networking-related activities.

This points to a high fit of generative AI for formalized, language-based tasks.

In summary, the results can be classified with respect to the research questions as follows:

- Regarding F1: At T01, significant differences in favor of administration are found for perceived usefulness and output quality. For perceived ease of use and reliability, no significant group differences emerge. At T02, the previously measured differences largely level out.
- Regarding F2: A significant increase in perceived usefulness and perceived ease of use is observed among scientific employees, whereas changes in administration are not statistically significant.
- Regarding F3: M365 Copilot is perceived as helpful in particular for structured, text-based activities (e.g., information search, analysis, text creation), whereas markedly lower usefulness ratings are found for learning and networking activities.

## 5 Discussion

The present study examined the introduction of M365 Copilot in a non-university research organization using a repeated cross-sectional survey (T01: n = 106; T02: n = 90). The focus was on the three research questions F1 to F3, addressing group-specific differences between scientific and administrative areas of work, temporal developments in acceptance characteristics, and task-related usefulness assessments. The sample was drawn as a quota sample from 550 licensed employees, yielding an organization-proximate but not representative picture. Analyses compared administration and science, taking into account method-inherent limitations due to self-selection, subjective self-reports, and the early implementation period.

Empirically, Copilot is evaluated predominantly positively across both measurement points, but with clear differences between administration and science. Descriptively, administration reports higher mean values; inferentially, significant group differences are found primarily for usefulness and output quality at T01, whereas no significant group differences emerge for perceived ease of use and reliability. Scientific employees start with more reserved, partly neutral assessments, but show marked increases over time, especially regarding productivity, effectiveness, and workload reduction.

Administration shows a high initial level and only moderate increases, indicating a saturation or plateau effect. Overall, operability is assessed as low-threshold and technical reliability as satisfactory. Notably, the assessment of technical reliability changes little over time and shows no significant group differences. This suggests that acceptance dynamics are explained primarily by work-related usefulness attributions rather than by technical stability aspects. Reported effect sizes are predominantly in the small to moderate range, indicating gradual adjustment processes and incremental learning trajectories rather than disruptive shifts in work perceptions.

Particularly noteworthy is the convergence of assessments between science and administration over time. The initially higher ratings in administration are relativized as scientific employees record significant gains. This points to learning and routinization effects rather than structurally different acceptance attitudes.

The temporal development indicates pronounced learning and habituation effects, particularly among scientific employees. For example, the perceived comprehensibility of interaction with Copilot ("clear interaction") improves markedly over time, whereas administration shows relatively stable, positive assessments early on. It is striking that usage aspects (output quality, clarity, operating comfort) remain predominantly positive but change far less than outcome- and task-related benefit indicators such as effectiveness, productivity, workload reduction, or faster task completion. This suggests that it is not so much the technical quality or usability of the system that increases over time, but rather users' ability to integrate Copilot purposefully into tasks where it is experienced as particularly helpful. With growing usage experience, sub-tasks that can be automated well are apparently identified; thus, perceived benefits increase even though the system characteristics themselves remain relatively constant.

Equally noteworthy is the consistently low assessment of potential harm from incorrect output. Ratings remain stable in the negative range across both groups and measurement points, indicating pronounced basic trust in controllable use or established verification and quality assurance routines.

The findings are consistent with core TAM assumptions in that not perceived ease of use alone, but particularly perceived usefulness is decisive for acceptance processes.

The task-related patterns can also be interpreted in terms of the task-technology-fit approach: high fit is found especially for clearly structured, language- and information-based tasks, while complex, creative, or socially embedded tasks benefit less to date.

This interpretation is supported by the qualitative study by CSIRO (Bano et al., 2025), based on interviews with 27 participants in a six-month M365 Copilot trial. However, Bano et al. (2025) point to the productivity paradox: automated time gains can be offset by correction and verification effort, especially for complex, domain-specific, or creative tasks, which occur more frequently among scientific employees.

From these findings, the following practical implications for introducing generative AI can be derived: context-sensitive implementation, work-context-specific training, capability building, and accompanying evaluations are necessary to identify usage barriers early. However, it should be noted that the positive ratings reported are based on subjective assessments; objective productivity measures are not available.

## 6 Limitations and outlook

As with all exploratory studies, this study is subject to several limitations. For example, a self-selection effect is likely, as more interested or tool-using persons are more likely to participate. Data collection is also based on self-reported assessments that may be subject to subjective biases and do not constitute an objective assessment. In addition, several central constructs (including output quality, reliability, harm from incorrect output, and voluntariness) were measured as single items or very short scales; measurement precision is therefore limited, and group and time differences derived from them (especially F1/F2) should be interpreted cautiously. Furthermore, the survey was conducted in a specific organizational and national environment, which limits transferability to other research organizations or international contexts—particularly given cultural and legal differences in dealing with AI and data protection. Convergent findings with the qualitative study by Bano et al. (2025) suggest a consistent tendency, but they do not replace representative validation. The very high internal consistency of the PU scale ($\alpha > .95$) could indicate content redundancy of items; future studies should test whether a shortened scale yields comparable results.

These limitations constrain the generalizability of the findings for answering research questions F1 to F3. In particular, it should be emphasized that observed group and time differences (F1, F2) are based on a non-representative, self-selected sample and reflect subjective assessments only. Likewise, task-related usefulness assessments (F3) can be generalized only to a limited extent. Sample sizes are also limited, especially for group-specific time comparisons, so that smaller effects may not have been detected statistically. Non-significant findings should therefore not necessarily be interpreted as an absence of an effect.

The task-related single items were intentionally analyzed in an exploratory descriptive manner in order to avoid alpha-error inflation due to numerous single comparisons. The patterns identified there should therefore be interpreted as initial indications and require targeted testing in future studies.

Although the key constructs were tested inferentially, it must be considered that multiple individual tests were conducted. No formal correction for multiple testing was applied in the exploratory design; the results should therefore primarily be understood as hypothesis-generating.

Because the data consist of two repeated cross-sectional samples in an early stage of implementation, observed time differences cannot unambiguously be interpreted as intra-individual learning processes. Changes between T01 and T02 reflect differences at the sample level and may be partly due to different compositions of participants. In addition, long-term effects on work processes, collaboration, and organizational structures could not be captured. Future research should therefore employ longitudinal designs and complement quantitative analyses with qualitative studies to systematically trace individual learning trajectories as well as organizational adaptation processes.

Due to the chosen approach and early survey period, medium- to long-term effects that might be expected—such as lack of transparency, emerging bias, insufficient accountability, or potential de-skilling effects—cannot be demonstrated. Especially for a research organization, these aspects are highly relevant and should therefore be addressed systematically in future quantitative surveys. For example, Krügel et al. (2025) show that users may be willing to integrate AI tools into their work but do not always have the ability to critically assess the quality and provenance of AI-generated content, underscoring the need to promote ethical education alongside digital competence.

Existing empirical results are so far still largely limited to single-case studies or small experimental comparisons; robust, controlled long-term studies with representative samples are lacking. Especially relevant for research institutions is the question of the role of AI in fostering creativity and innovation capability and its impacts on collaboration in research teams. AI-implicated social effects such as willingness and effectiveness to collaborate are still under-researched.

The present results highlight the importance of socio-organizational support for introducing generative AI in addition to technical enablement. The differing assessments of administration and science illustrate how strongly perception and use of M365 Copilot are shaped by task profiles, experience levels, and institutional conditions. Sustainable success therefore depends not on the availability of powerful models alone, but on the systematic design of implementation processes, capability building, and governance structures.

The results suggest that generative AI can realize its potential particularly when employees are involved early and supported in developing appropriate usage patterns that address efficiency gains as well as questions of quality, responsibility, and ethics. The findings support viewing generative AI less as a replacement for human expertise and more as a complementary tool, whose added value unfolds especially in the combination of technical capability with domain expertise and organizational embedding.

## Author addresses

Carsten F. Schmidt
Fraunhofer-Institut für Arbeitswirtschaft und Organisation IAO
Nobelstr. 12
D-70569 Stuttgart
carsten.schmidt@iao.fraunhofer.de
Dr. Sophie Petzolt
Fraunhofer-Institut für Arbeitswirtschaft und Organisation IAO
Nobelstr. 12
D-70569 Stuttgart
sophie.petzolt@iao.fraunhofer.de
Dr. Wolfgang Beinhauer
Fraunhofer-Institut für Arbeitswirtschaft und Organisation IAO
Nobelstr. 12
D-70569 Stuttgart
wolfgang.beinhauer@iao.fraunhofer.de
Prof. Dr. Ingo Weber
Fraunhofer-Gesellschaft zur Förderung der angewandten Forschung e.V.
Hansastraße 27 c
D-80686 München
ingo.weber@zv.fraunhofer.de
Stefan Langer
Fraunhofer-Gesellschaft zur Förderung der angewandten Forschung e.V.
Hansastraße 27 c
D-80686 München
stefan.langer@zv.fraunhofer.de